\documentclass{appolb}
\usepackage{epsfig}
\usepackage{graphicx}

\def\bc{\begin{center}}
\def\ec{\end{center}}
\def\beq{\begin{equation}}
\def\eeq{\end{equation}}
\def\bs{\begin{slide}}
\def\es{\end{slide}}
\newcommand{\bmath}{\begin{displaymath}}
\newcommand{\emath}{\end{displaymath}}
\newcommand{\beqn}{\begin{eqnarray}}
\newcommand{\eeqn}{\end{eqnarray}}
\newcommand{\beqns}{\begin{eqnarray*}}
\newcommand{\eeqns}{\end{eqnarray*}}
\newcommand{\ba}{\begin{array}{c}} 
\newcommand{\bat}{\begin{array}{cc}} 
\newcommand{\ea}{\end{array}}

\begin{document}
\title{Large-$N_C$ estimate of the chiral low-energy constants \footnote{IFIC/07-10 report. Talk given
at the Final Euridice Meeting, 24th-27th August 2006, Kazimierz (Poland).}
}
\author{J. Portol\'es
\address{Instituto de F\'{\i}sica Corpuscular, IFIC, CSIC-Universitat de Val\`encia, \\ Apt. Correus 22085,
E-46071 Val\`encia, Spain}
}
\maketitle
\begin{abstract}
Chiral low-energy constants incorporate short-distance information from the dynamics involving heavier degrees of
freedom not present in the chiral Lagrangian. We have studied the contribution of the lightest resonances
to the chiral low-energy constants, up to ${\cal O}(p^6)$, within a systematic procedure guided by the 
large-$N_C$ limit of QCD and also including short-distance asymptotic constraints. 
\end{abstract}
  
\section{Introduction}
Chiral symmetry of massless Quantum Chromodynamics has become the key tool in the study of the very low-energy domain
of strong interactions (typically $E \sim M_{\pi}$), where QCD turns non-perturbative. It is indeed the guiding principle
in the construction of Chiral Perturbation Theory ($\chi PT$), the effective field theory of QCD in this energy region
\cite{Weinberg:1978kz,Gasser:1983yg,Gasser:1984gg}. The $\chi PT$ Lagrangian has a perturbative structure guided
 by powers of external
momenta and light quark masses. It involves the multiplet of pseudoGoldstone bosons, i.e. pseudoscalar mesons ($\pi$,
$K$, $\eta$), and classical auxiliary fields. The theory, up to a fixed order in the expansion ${\cal O}(p^n)$, can be
obtained by a construction guided by chiral symmetry~:
\begin{equation}
{\cal L}_{\chi PT} \, = \, {\cal L}_2^{\chi PT} + {\cal L}_4^{\chi PT} + {\cal L}_6^{\chi PT} + ... + 
{\cal L}_n^{\chi PT} \, .
\end{equation}
${\cal L}_2^{\chi PT}$ embodies the spontaneous breaking of the chiral symmetry and depends only on two parameters~: $F$,
the decay constant of the pion, and $B_0 F^2 = - \langle 0 | \overline{\psi} \psi |0 \rangle$, the vacuum expectation
value of the light quarks; both of them in the chiral limit. Higher orders in the expansion bring in the information
from short-distance contributions that have been integrated out, for instance resonance states. As in any effective field
theory (EFT) this information is incorporated into the low-energy constants (LECs) that weight the operators of the theory~:
\begin{eqnarray} \label{eq:chpt}
{\cal L}_4^{\chi PT} \, = \, \sum_{i=1}^{10} L_i {\cal O}_i^{(4)} \, , & \; \; \; \; \; \; \; \; \; \; \; \; 
\; \; \;  \; \; \; & \; 
{\cal L}_6^{\chi PT} \, = \, \sum_{i=1}^{90} C_i {\cal O}_i^{(6)} \; , 
\end{eqnarray}
for $SU(3)$. Explicit expressions for the operators can be read from Refs.~\cite{Gasser:1984gg,Bijnens:1999sh}. The scale
that specifies the chiral expansion, $\Lambda_{\chi} \sim M_V$ (being $M_V$ the mass of the $\rho(770)$, the lightest
hadron not included in the theory), indicates that LECs in $\chi PT$ should receive
contributions from the energy regime at or above that scale \cite{Georgi:1991ch}. The determination of the contributions
of the lightest multiplets of resonances to the ${\cal O}(p^4)$ LECs in ${\cal L}_4^{\chi PT}$ \cite{Ecker:1988te} has
shown that they indeed saturate the values extracted from the phenomenological analyses. As a consequence it is reasonable to
think that the most important contribution to the LECs is provided by the energy region immediately above the integrated
scale ($E \sim \Lambda_{\chi}$).

\section{The role of Resonance Chiral Theory}
As illustrated in the ${\cal O}(p^4)$ case \cite{Ecker:1988te,Ecker:1989yg} a procedure to systematically determine
the resonance contributions to the LECs in $\chi PT$ is available. Essentially the idea is to construct a Lagrangian
theory in terms of resonances, pseudoscalar mesons and auxiliary fields respecting the underlying chiral symmetry. Then,
upon integration of the heavier states, the $\chi PT$ Lagrangian is recovered. The outcome of this first step is that
LECs are traded by the equally unknown couplings of the resonance Lagrangian though, at this point, one may also notice
relations between LECs. In a second stage information on the resonance couplings is obtained, whether from phenomenology
or, more interestingly, by imposing theoretical constraints from the QCD asymptotic behaviour of form factors or Green 
functions. 
\par
Contrarily to $\chi PT$, the lack of a mass gap between the spectrum of light-flavoured resonances and the perturbative
continuum prevents the construction of an appropriate EFT to describe the interaction of resonances
and pseudoscalar mesons. However there are several tools that allow us to grasp relevant features of QCD and to 
implement them in an EFT-like Lagrangian model. The two relevant basis are the following~:
\vspace*{0.2cm} \\
i) A theorem put forward by S.~Weinberg \cite{Weinberg:1978kz} and worked out by H.~Leutwyler \cite{Leutwyler:1993iq}
states that if one writes down the most general possible Lagrangian, including all terms consistent with  
assumed symmetry principles, and then calculates matrix elements with this Lagrangian to any given order of
perturbation theory, the result will be the most general possible S--matrix amplitude consistent with analyticity, 
perturbative unitarity, cluster decomposition and the principles of symmetry that have been required. 
\vspace*{0.1cm} \\
ii) The inverse of the number of colours of the $SU(N_C)$ gauge group can be taken as a perturbative expansion
parameter \cite{'tHooft:1974hx}. Large-$N_C$ QCD shows features that resemble, both qualitatively and quantitatively,
the $N_C=3$ case \cite{Peris:1998nj}. In practice, the consequences of this approach are that meson dynamics in the
large-$N_C$ limit
is described by tree diagrams of an effective local Lagrangian involving an infinite spectrum of zero-width mesons.
\vspace*{0.2cm} 
\par
Both statements can be combined by constructing a Lagrangian theory in terms of $SU(3)$ (pseudoGoldstone mesons) 
and $U(3)$ (heavier resonances) flavour multiplets as active degrees of freedom. This has been established
\cite{Ecker:1988te,Ecker:1989yg,Cirigliano:2006hb} systematically and devises what is known as Resonance Chiral
Theory ($R \chi T$) that shows the following main features~: \vspace*{0.2cm} \\
1) The construction of the operators in the Lagrangian is guided by chiral symmetry for the pseudoGoldstone
mesons and by unitary symmetry for the resonances. The general structure of these couplings is~:
\begin{equation} \label{eq:structure}
{\cal O} \, = \, \langle \, R_1 R_2 ...R_m \, \chi(p^n)  \, \rangle \, \; \; \; \in \; \; \; \, 
{\cal L}_{(n)}^{\overbrace{\scriptstyle RR...R}^{m}} \; ,
\end{equation}
where  $R_j$ indicates a resonance field and $\chi (p^n)$ is a chiral structured tensor, involving the pseudoGoldstone
mesons and auxiliary fields only. Then, the usual chiral counting in $\chi PT$ \cite{Weinberg:1978kz} represented by the 
power of momenta can straightforwardly be applied to $\chi (p^n)$. With these settings chiral symmetry is preserved
upon integration of the resonance fields and, at the same time, the low--energy behaviour of the amplitudes is 
guaranteed.  
\vspace*{0.1cm} \\
2) Symmetries do not provide information on the coupling constants as these incorporate short-distance
dynamics not included explicitly in the Lagrangian. The latter 
is supposed to bridge between the energy region below resonances ($E \ll M_V$) and the parton
regime ($E \gg M_V$). This hypothesis indicates that it should match both regions and it satisfies,
by construction, the chiral constraints. To suit the high-energy behaviour one can match, for instance,
the OPE of Green functions (that are order parameters of chiral symmetry) with the corresponding expressions evaluated
within our theory. In addition the asymptotic trend of form factors of QCD currents is estimated  from the 
spectral structure of two-point functions or the partonic make-up and it is enforced on the couplings. 
This heuristic strategy is well supported by the
phenomenology \cite{Peris:1998nj,Cirigliano:2006hb,Knecht:2001xc,Cirigliano:2004ue,Cirigliano:2005xn}.
\vspace*{0.2cm}
\par
With this pattern the content of the theory is, schematically, given by~:
\begin{equation}
{\cal L}_{R \chi T} \, = \, {\cal L}_2^{\chi PT} \, + \, \sum_n \, {\cal L}_{n>2}^{GB} \, + {\cal L}_R \; , 
\end{equation}
where ${\cal L}_{n>2}^{GB}$ has the same structure than ${\cal L}_4^{\chi PT}$, ${\cal L}_6^{\chi PT}$, ...
in Eq.~(\ref{eq:chpt})
though with different coupling constants, and ${\cal L}_R$ involves terms with resonances and their couplings
to pseudoGoldstone modes. 
\par
$R \chi T$ lacks an expansion parameter. There is of course the guide provided by $1/N_C$ that translates into the
loop expansion, however there is no counting that limits the number of operators with resonances that have to be 
included in the initial Lagrangian. However the number of resonance fields to be kept relies fundamentally in the physical 
system that we are interested in and the maximum order of the chiral tensor $\chi (p^n)$ 
in Eq.~({\ref{eq:structure}) is very much constrained by the required high--energy behaviour. 
\par
As commented above large-$N_C$ requires, already at $N_C \rightarrow \infty$, an infinite spectrum in order to match 
the leading QCD logarithms, though we do not know how to implement this in a model-independent way. The usual approach
in $R \chi T$ is to include the lightest resonances because their phenomenological relevance, though there is no
conceptual problem that prevents the addition of a finite number of multiplets. This cut in the spectrum  may produce
inconsistencies in the matching procedure outlined above \cite{Bijnens:2003rc}. To deal with this one can include
more states that may delay the appearance of that problem. 

\section{Resonance contributions to the ${\cal O}(p^6)$ LECs}
In Ref.~\cite{Cirigliano:2006hb} we have constructed the $R \chi T$ Lagrangian needed to evaluate the resonance
contributions to the ${\cal O}(p^6)$ LECs in Eq.~(\ref{eq:chpt}). It has the following structure~:
\begin{equation} \label{eq:rcht}
{\cal L}_{R\chi T} \, = \, {\cal L}_2^{\chi PT}  +  {\cal L}_4^{GB}  +  {\cal L}_6^{GB}  +  
{\cal L}_{\mathrm{kin}}^R    + {\cal L}_{(2)}^R  +  {\cal L}_{(4)}^R  +  {\cal L}_{(2)}^{RR}  +
  {\cal L}_{(0)}^{RRR}  + 
{\cal L}_{\varepsilon} \; .
\end{equation}
The inclusion of spin-1 resonances has been performed within the antisymmetric tensor formalism. However there are
independent contributions to the ${\cal O}(p^6)$ LECs coming from odd-intrinsic-parity couplings involving spin-1 
resonances in the Proca formalism. These are included in ${\cal L}_{\varepsilon}$. In the above representation
it can be shown \cite{Ecker:1989yg} that, at  ${\cal O}(p^4)$, all local terms in ${\cal L}_4^{GB}$ have to vanish
in order not to upset the asymptotic behaviour of QCD correlators. A corresponding result at ${\cal O}(p^6)$ is still
lacking but we have also assumed that all the couplings in ${\cal L}_6^{GB}$ are set to zero.
\par
${\cal L}_{R \chi T}$ in Eq.~(\ref{eq:rcht}) involves 6 a priori unknown couplings in ${\cal L}_{(2)}^R$, 70 in 
${\cal L}_{(4)}^R,$ 38 in ${\cal L}_{(2)}^{RR}$, 7 in ${\cal L}_{(0)}^{RRR}$ and 3 in ${\cal L}_{\varepsilon}$. Some
additional work provides an enormous simplification~: \vspace*{0.2cm} \\
i) Upon integration of resonances not all couplings appear independently in the LECs. In general only several 
combination of couplings intervene and to take into account this case one can perform suitable redefinitions of
the fields. This procedure
spoils in general the high-energy behaviour of the theory but it is correct for the evaluation of the LECs. Indeed
the 70 couplings in ${\cal L}_{(4)}^R$ reduce to 23. \vspace*{0.1cm} \\
ii) 
The next step is to enforce additional short-distance information, i.e. the leading behaviour at large momenta, for 
two and three-point functions and form factors. This procedure, set in Ref.~\cite{Ecker:1989yg}, relies in well-known
features of partonic scattering or asymptotic QCD \cite{Froissart:1961ux}. 
Two-current correlators and associated form-factors provide 19 new constraints on couplings, while the three-point
Green functions studied till now~: $\langle VAP \rangle$ \cite{Cirigliano:2004ue} and $\langle SPP \rangle$
\cite{Cirigliano:2006hb,Cirigliano:2005xn}, give 6 and 5 independent restrictions, respectively. 
\vspace*{0.2cm} \\
\hspace*{0.5cm} Hence we can already determine
fully the resonance contribution to the ${\cal O}(p^6)$ couplings $C_{78}$ and $C_{89}$ (that appear in $\pi \rightarrow 
\ell \nu_{\ell} \gamma$ and $\pi \rightarrow \ell \nu_{\ell} \gamma^*$, respectively), $C_{87}$ (in 
$\langle A_{\mu}  A_{\nu} \rangle$), $C_{88}$ and $C_{90}$ (in $F_V^{\pi}(q^2)$ and the $q^2$ dependence of the
form factors in $K_{\ell 3}$), $C_{38}$ (in
$\langle SS \rangle$) and $C_{12}$ and $C_{34}$ (in $F_V^{\pi,K}(q^2)$ and $f_+^{K^0 \pi^-}(0)$). 
\par
It is significant to observe that, as shown in the analysis of the $\langle SPP \rangle$ Green function, the use of 
a Lagrangian theory like $R \chi T$ though involved \cite{Cirigliano:2006hb} brings more information (encoded
in the symmetries of the Lagrangian) than the use of a parametric ansatz \cite{Cirigliano:2005xn}.

\section{$K_{\ell 3}$ decays~: determination of $f_+^{K^0 \pi^-}(0)$}
$K_{\ell 3}$ decays have the potential to provide one of the most accurate determinations of the $V_{us}$ CKM
element.  The main uncertainty in extracting this parameter comes from theoretical calculations of the
vector form factor $f_{+}^{K^0 \pi^-}(0)$  defined by~:
\begin{equation}
\langle \, \pi^-(p) \, | \,  \overline{s} \gamma_{\mu} u \, | \, K^0(q) \, \rangle \, = \, f_{+}^{K^0 \pi^-}(t) \, 
 (q+p)_{\mu} \, + \, 
f_{-}^{K^0 \pi^-}(t) \,  (q-p)_{\mu} \, , 
\end{equation}
with $t=(q-p)^2$. Deviations of $f_{+}^{K^0 \pi^-}(0)$ from unity (the octet symmetry limit) are of second order
in the $SU(3)$ breaking~: 
\begin{equation}
 f_{+}^{K^0 \pi^-}(0) \, = \, 1 \, + \, f_{p^4} \, + \, f_{p^6} \, + ...
\end{equation}
The first correction is ${\cal O}(p^4)$ in $\chi PT$, through one loop calculation and no local terms,
and it gives $f_{p^4} = -0.0227$, essentially without uncertainty \cite{Leutwyler:1984je}. At ${\cal O}(p^6)$
two-loop, one-loop and local terms contribute and the latter make the determination more uncertain. Loops give
$f_{p^6}^{\mathrm{loops}}(M_{\rho}) = 0.0093(5)$~\cite{Cirigliano:2005xn,Bijnens:2003uy}. The explicit form for
the tree-level contribution is~:
\begin{equation} \label{eq:tree}
f_{p^6}^{\mathrm{tree}}(M_{\rho}) \, = \, 8 \frac{\left( M_K^2 - M_{\pi}^2 \right)^2}{F_{\pi}^2} \, 
\left[ \, \frac{\left( L_5^r(M_{\rho}^2) \right)^2}{F_{\pi}^2} - C_{12}^r(M_{\rho}^2) - C_{34}^r(M_{\rho}^2) \, 
\right] \; ,
\end{equation}
that involves LECs of $\chi PT$ both at ${\cal O}(p^4)$ and ${\cal O}(p^6)$. 
In Ref.~\cite{Cirigliano:2005xn}, and using the method outlined in these proceedings, we have determined the
contribution of scalar and pseudoscalar resonances to the LECs present in Eq.~(\ref{eq:tree}), and we get~:
\begin{equation}
f_{p^6}^{\mathrm{tree}}(M_{\rho}) \, = \, - \frac{\left( M_K^2 - M_{\pi}^2 \right)^2}{2 M_S^4} \left( 1-
\frac{M_S^2}{M_P^2} \right)^2\; .
\end{equation}

\begin{figure}[!t]
\centering

\begin{picture}(300,175)  
\put(140,65){\makebox(50,50){\epsfig{figure=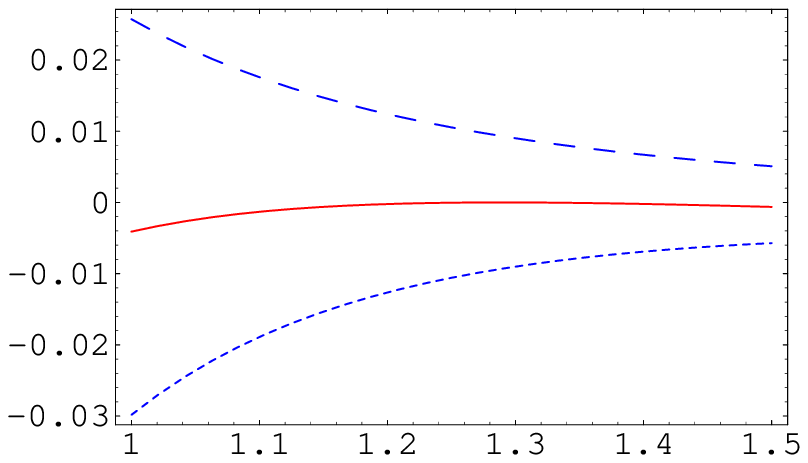,width=9.5cm}}}
\put(240,0){
{\large 
$M_S \ ({\rm GeV})$ 
}
}
\put(-11,160){
{\large 
$ f_{p^6}^{\rm tree} (M_\rho)  $ 
}
}
\put(150,140){
{\small
$L_5 \times L_5 / F_\pi^2 $
}
}
\put(140,50){
{\small
$- (C_{12} + C_{34})$
}
}
\end{picture}
\caption{
\label{fig:ftree}
We display $f_{p^6}^{\rm tree} (M_\rho)$ as a function of $M_S$ for $M_P=1.3$ GeV 
(solid line). 
We also plot the two components: the dashed line represents  the 
term proportional to $L_5 \times L_5$, while the 
dotted line represents the term proportional to 
$- (C_{12} + C_{34})$. The cancellation between both contributions is very large. 
}
\end{figure}

As can be seen in Figure~1, it produces a tiny result~: $f_{p^6}^{\mathrm{tree}}(M_{\rho}) = -0.002(12)$ due to a strong
cancellation between both terms. We end up with the
final result~:
\begin{equation}
f_{+}^{K^0 \pi^-}(0) \, = \, 0.984(12) \; .
\end{equation}

\begin{table}
\begin{center}
\begin{tabular}{|l|l|}
\hline
\multicolumn{1}{|c}{Reference} &
\multicolumn{1}{|c|}{$f_{+}^{K^0 \pi^-}(0)$} \\
\hline
Quark model \cite{Leutwyler:1984je} &  $0.961(8)$ \\
Lattice (quenched) \cite{Becirevic:2004ya} & $0.960(9)$ \\
Lattice (unquenched) \cite{Dawson:2006qc}& $0.968(11)$ \\
Lattice (unquenched) \cite{Zanotti}  & $0.961(5)$ \\
(Chiral +  \cite{Leutwyler:1984je}) \cite{Bijnens:2003uy} & $0.971(10)$ \\
$K \pi$ scalar f.f. \cite{Jamin:2004re} & $0.974(11)$ \\
Ours \cite{Cirigliano:2005xn} & $0.984(12)$ \\
\hline
\end{tabular} 
\end{center}
\caption{Comparison of different predictions for $f_{+}^{K^0 \pi^-}(0)$. The value
quoted for Ref.~\cite{Bijnens:2003uy} has been modified as explained in 
Ref.~\cite{Cirigliano:2005xn}.}
\end{table}

In Table~1 we compare our result with other determinations coming from different sources. 
This comparison shows a clear pattern~: both quenched and unquenched lattice results are in very good agreement with
the quark model prediction by Leutwyler and Roos \cite{Leutwyler:1984je}, while our analytic determination of 
$f_{+}^{K^0 \pi^-}(0)$ shows a clear tension. It is also 
interesting to notice the tiny modification that unquenching produces in the lattice results though, as shown in 
Ref.~\cite{Bijnens:2003uy}, chiral logarithms are very much important in the determination of $f_{+}^{K^0 \pi^-}(0)$.
A better understanding is still required on this issue.

\section{Perspective}
The study of the resonance energy region is essential
for the understanding of hadron phenomenology driven by non-perturbative QCD~: hadronic tau decays, final-state
interactions, kaon decays, etc.
\par
$R \chi T$ is a systematic setting that allows to implement known features of QCD into a Lagrangian framework. We have
shown that it provides sensible results but much more work is still needed to reach a better understanding both on the
phenomenology and on the underlying ideas that joint together in this formulation. 

\section*{Acknowledgements}
I wish to thank M.~Krawczyk, G.~Pancheri and H.~Czyz for the excellent organization of the Final Euridice meeting.
This work has been supported in part by HPRN-CT-2002-00311 (EURIDICE), by the EU MRTN-CT-2006-035482 (FLAVIAnet), 
by MEC (Spain) under grant FPA2004-00996 and
by Generalitat Valenciana under grants ACOMP06/098 and GV05/015.

\end{document}